\documentclass[
pra,superscriptaddress,showpacs]{revtex4}
\usepackage{amsbsy}
\usepackage{amsmath}
\usepackage{amsfonts}
\usepackage{graphicx}
\usepackage{color}
\usepackage{epsfig}

\newcommand{\dd}{\,{\rm d}}
\newcommand{\ii}{{\rm i}}
\newcommand{\A}{A}

\newtheorem{definition}{Definition}
\newtheorem{lemma}{Lemma}
\newtheorem{theorem}{Theorem}
\newtheorem{identity}{Identity}

\newtheorem{conjecture}{Conjecture}

\errorstopmode

\begin{document}

\title{Operational formulation of homodyne detection}
\author{Tom\'a\v s Tyc}
\affiliation{Institute of Theoretical Physics, Masaryk University, 61137 Brno,
Czech Republic\footnote{Major part of this work has been done at Department of
Physics and Centre for Advanced Computing --
Algorithms and Cryptography, Macquarie University Sydney, New
South Wales 2109, Australia }}
\author{Barry C.~Sanders}
\affiliation{Institute for Quantum Information Science, University of
        Calgary, Alberta T2N 1N4, Canada}
\affiliation{Australian Centre for Quantum Computer Technology, Macquarie
        University, Sydney, New South Wales 2109, Australia}
\date{July 8, 2004}

\begin{abstract} 
 We obtain the standard quadrature-phase positive operator-valued measure
 (POVM) for homodyne detection directly and rigorously from the POVM for 
 photon counting without directly employing the mean field approximation for
 the local oscillator. In addition we obtain correction terms for the
 quadrature-phase POVM that are applicable for relatively weak local oscillator
 field strengths and typical signal states.
\end{abstract}
\maketitle

\section{Introduction}
\label{sec:Intro}

With the advent of squeezed states of light~\cite{Rad87}, a full quantum
description of optical homodyne detection~\cite{Rad87,Yue78,Yue79,Yue83}
assumed importance as homodyne detection (HD) yields phase--dependent
measurements of the light field.  Whereas photodetectors acquire
phase--insensitive information about photon statistics~\cite{Gla63,Sud63},
homodyne detection mixes the signal field with a coherent local oscillator (LO)
to yield photon statistics on the output fields that depend on the
phase~$\varphi$ of the LO.  By varying this phase~$\varphi$, phase--dependent
properties of the signal state~$\hat\rho$ can be inferred.  The
phase--sensitive measurement with respect to the in--phase quadrature~$x$ or
its canonically conjugate out--of--phase quadrature~$p$, or some in--between
quadrature $x_\varphi \equiv p_{\varphi-\pi/2}\equiv
x\cos\varphi+p\sin\varphi$, is necessary to observe the nonclassical properties
of squeezed light.  Phase--sensitive measurement has developed beyond measuring
specific quadrature--phase statistics to acquiring information for many values
of~$\varphi$ and reconstructing the density matrix~$\hat\rho$ for the signal
field.  This technique, known as optical homodyne tomography~\cite{Ris},
illustrates another important application of homodyne detection.  Homodyne
detection has evolved into a key tool of quantum optics with applications
including squeezed light detection, optical homodyne tomography and continuous
variable quantum teleportation \cite{Vai94,Bra98,Fur98}.

\begin{figure}[h]
\includegraphics*[width=3.5cm,keepaspectratio]{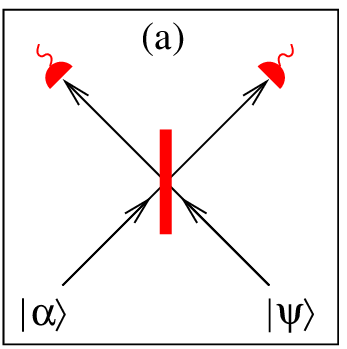} \hspace{2cm}
\includegraphics*[width=3.5cm,keepaspectratio]{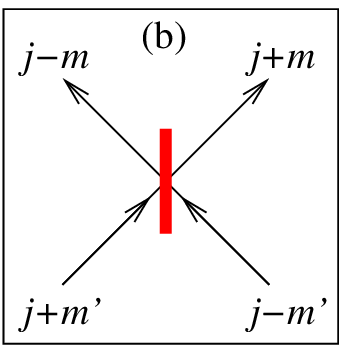}
\caption{Balanced homodyne detection scheme: (a)~the input state
$|\psi\rangle$ is mixed with a LO in coherent state~$|\alpha\rangle$, and
photon counting occurs at the two output ports; (b)~photon numbers $j \pm m'$
are shown entering the two input ports and $j \pm m$ are counted at the
output.}
\label{4PHD}
\end{figure}

The homodyne detection scheme discussed above involves mixing the signal field
with a LO field at a beamsplitter (BS), and the two output fields are subjected
to photodetection, as shown in Fig.~\ref{4PHD}.  The measured photodetection
statistics are analyzed to infer the quadrature--phase statistics.  Only in the
limit of infinite LO field strengths can the measurement be said to correspond
to quadrature--phase measurements, and, of course, this limit is in principle
unattainable.  However, a good approximation to quadrature--phase measurements
is attained. In the most useful variant, a 50/50 BS is used, and the difference
between the photocounts at the two output ports is used to infer the
quadrature--phase statistics.  This is known as balanced homodyne detection
(BHD) and has the advantage of automatically canceling the photon number sum
at the two input ports from the detected output fields.

The description of homodyne detection begins with photodetection of the output
fields and then, to validate the approximations normally applied in homodyne
detection of quadrature--phase POVM, must show that the resultant two--mode
photon statistic reduces in some way to the quadrature--phase distribution, for
the signal field~$\hat\rho$.  This connection between photon statistics to
quadrature--phase, or joint quadrature--phase, measurements has been
established via calculations involving quasi-probability distributions or
characteristic functions (moment-generating functions) for the electromagnetic
field and allowing the local oscillator strength to become infinitely large.
Yuen and Shapiro introduced the characteristic function approach in their
seminal quantum theory of HD~\cite{Yue78}, and Walker employs Wigner functions
in his analysis of HD~\cite{Wal87}.  Braunstein~\cite{Bra90} uses the positive
$P$-representation in the description of the photon counting statistics, and he
emphasizes the quantum nature of the LO as he investigates ``the effects of a
finite--amplitude fully-quantum-mechanical local oscillator''.  Banaszek and
W\'{o}dkiewicz~\cite{Ban97} calculate moments of operationally defined
quadrature operators, with an emphasis on finite photodetection efficiency, but
in contrast to our approach, employ the mean field approximation to the LO from
the outset. 

These studies undoubtedly establish the connection between the exact
photodetection statistics and the approximate quadrature--phase HD.  However,
modern applications of homodyne detection, for example to quantum information
applications such as continuous--variable quantum teleportation, requires an
operational quantum theoretic approach~\cite{Rud01}; Banaszek and
W\'{o}dkiewicz advocate the operational approach, but here we avoid the mean
field approximation and thereby include correction terms for the POVM
corresponding to HD. The operational approach is important in the context that
a measurement may be applied for some purpose
other than characterizing the state~$\hat\rho$; paradoxically, in
continuous--variable quantum teleportation~\cite{Bra98,Fur98}, the sender mixes
the field described by density operator $\hat\rho$ with one component of a
two--mode squeezed vacuum state~\cite{Sch85} in such a way that the sender
cannot know, even in principle, what the density operator~$\hat\rho$ was that
is being subjected to this measurement.  For such applications, a rigorous
approach to homodyne measurement, which demonstrates that the POVM for
photodetection reduces to the POVM for quadrature--phase or joint
quadrature--phase measurements is necessary.  Here we establish this connection
between actual and convenient POVMs by directly calculating the photon counting
probabilities using two different approaches: (i) working in the Fock basis for
the Hilbert space of the signal and LO modes, we employ asymptotic expressions
for SU(2) Wigner functions that are the BS matrix elements in the Fock basis;
(ii) working in the over-complete basis of coherent states and taking advantage
of the simple transformation of coherent states at the BS, we employ the
Glauber-Sudarshan $P$-function.

\section{Balanced homodyne detection scheme}

A balanced homodyne detection scheme is depicted in Fig.~\ref{4PHD}(a). The
(generally mixed) signal state $\hat\rho$ to be measured is coherently mixed at
the BS with a LO assumed to be in a coherent state (in the optical domain a
coherent state with an absolute adjustable phase has not been achieved, but the
coherent state approach leads to correct measured results provided that the
signal field and LO field are derived from the same source~\cite{Mol97,Rud01}).
The photon number difference from the two BS output ports is measured.  The
photon number sum can also be measured but usually is not.  However, in our
analysis we include the treatment of both the difference and the sum as this is
a more complete description than considering the difference alone. We will
denote the photon number difference by $2m\in {\mathbb Z}$ and the sum by
$2j\in {\mathbb Z}$.

The Hilbert space of two modes of electromagnetic field has the basis
$\{|n_1\rangle\otimes|n_2\rangle\}$ of joint eigenstates of the photon number
operators $\hat n_1=\hat a_1^\dagger\hat a_1$ and $\hat n_2=\hat
a_2^\dagger\hat a_2$. Denoting $j=(n_1+n_2)/2$, $m=(n_1-n_2)/2$, we will use
the notation $|jm\rangle\equiv|n_1\rangle\otimes|n_2\rangle$. Thus, the state
$|jm\rangle$ is the number states with photon numbers $j\pm m$ at modes 1 and
2, respectively. The value of $j$ can be any non-negative half-integer and $m$
can get values $-j,-j+1,\dots,j$ for a given $j$.

\subsection{Beam splitter transformation}

The beam splitter action on a two-mode state of electromagnetic field is
given by the SU(2) transformation \cite{Yur86,Cam89,San95}
\begin{equation}
   \hat B(\varphi_1,\varphi_2,\varphi_3)={\rm e}^{-i\varphi_1\hat J_z}
{\rm e}^{-i\varphi_2\hat J_y}{\rm e}^{-i\varphi_3\hat J_z},
\end{equation}
where the SU(2) generators $\hat J_y,\hat J_z$ are expressed in the
Schwinger boson representation as  
\begin{equation}
  \hat J_y=-\frac{\rm i}2(\hat a_1^\dagger\hat a_2-\hat a_2^\dagger\hat a_1),
 \quad \hat J_z=\frac12(\hat a_1^\dagger\hat a_1-\hat a_2^\dagger\hat a_2)
\end{equation}
An input state $|jm\rangle$ is transformed under the BS action as
\begin{equation}
 \hat B(\varphi_1,\varphi_2,\varphi_3)|jm\rangle
  =\sum_{m'}{\rm e}^{-i(m\varphi_3+m'\varphi_1)}
                d^j_{m'm}(\varphi_2)\,|jm'\rangle,
\end{equation}
where $m'$ in the sum runs from $-j$ to $j$ with unit steps and
$d^j_{m'm}(\varphi_2)=\langle jm'|{\rm e}^{-i\varphi_2\hat J_y}|jm\rangle$ are
the SU(2) Wigner functions \cite{Row01}.  

On the other hand, coherent states are transformed in a very simple way on BS.
If the initial two-mode coherent state is
$|\alpha_1\rangle\otimes|\alpha_2\rangle$, where
\begin{equation}
  |\alpha_i\rangle ={\rm e}^{-|\alpha_i|^2/2}\sum_{n=0}^\infty\,
  \frac{\alpha_i^n}{\sqrt{n!}}\,|n\rangle,
\end{equation}
then the BS output state is again a two-mode coherent state with amplitudes
$\alpha_1',\alpha_2'$:
\begin{equation}
 \hat B(\varphi_1,\varphi_2,\varphi_3)|\alpha_1\rangle\otimes|\alpha_2\rangle
=|\alpha_1'\rangle\otimes|\alpha_2'\rangle
\end{equation}
with $\alpha_1'={\rm e}^{-{\rm i}\varphi_1/2}(\alpha_1\cos\varphi_2{\rm
e}^{-{\rm i}\varphi_3 /2}-\alpha_2\sin\varphi_2{\rm e}^{{\rm i}\varphi_3/2})$
and $\alpha_2'={\rm e}^{{\rm i}\varphi_1/2}(\alpha_1\sin\varphi_2{\rm e}^{-{\rm
i}\varphi_3 /2}+\alpha_2\cos\varphi_2{\rm e}^{{\rm i}\varphi_3/2})$.  This
simple transformation is a key reason for the usefulness of the
Glauber-Sudarshan P function in describing homodyne detection.

For the rest of the paper, we will consider BHD with no phase factors, so we
set $\varphi_1=0,\varphi_2=\pi/2,\varphi_3=0$.  Ideally, the LO is prepared in
the pure coherent state with amplitude $\alpha$, and is directed into port 1 of
BS. The unknown signal field described by the density operator $\hat\rho$
enters the second input port.  The total state of the two modes before entering
BS is then
\begin{equation}
  \hat\rho_{\rm in}=|\alpha\rangle \langle\alpha|\otimes\hat\rho
\end{equation}
The beam splitter transforms the input state into
\begin{equation} 
  \hat\rho_{\rm out}=\hat B\hat\rho_{\rm in}\hat B^\dagger
\end{equation}
The probability of detecting $j+m$ and $j-m$ photons at the two BS outputs is
then 
\begin{equation}
 P^j_m={\rm Tr}(\hat\rho_{\rm out}|jm\rangle\langle jm|)
      =\langle jm|\hat B\hat\rho_{\rm in}\hat B^\dagger|jm\rangle
\label{1probab}\end{equation}
The probability $P^j_m$ can be expressed as
\begin{equation} 
 P^j_m={\rm Tr}(\hat\rho_{\rm out}\hat{\sf E}_m^j),
\end{equation}
where the POVM $\hat{\sf E}_m^j = |jm\rangle\langle jm|$ satisfies the
completeness condition $\sum_{m,j}\hat{\sf E}_m^j =\openone$ and positivity
condition ${\rm Tr}(\hat\rho_{\rm in}\hat{\sf E}_m^j)\ge0$.  The importance of
the photon number difference measurement and its relation to phase measurements
has been emphasized for many years, including in early work on phase operators
in two mode systems~\cite{Sus64}. If the total photon sum $2j$ is not measured
in BHD and only the difference $2m$ is observed, the appropriate POVM is
\begin{align}
\hat{\sf E}_m &= \sum_{j=|m|,|m|+1,\dots}^\infty |jm\rangle\langle jm|
                \nonumber       \\      &
       = \sum_{j=|m|,|m|+1,\dots}^\infty|j+m\rangle_1\langle j+m|\otimes
                |j-m\rangle_2\langle j-m|\,,
\label{no2j}\end{align}
where the subscripts refer to the output ports of BS.  However, we consider
here the more valuable case when both $m$ and $j$ are measured.

\section{Asymptotic SU(2) Wigner function approach}
\label{asymp}

In this section we derive the photon counting probability $P^j_m$ in the strong
LO limit using the asymptotic formul{\ae} for SU(2) Wigner functions.  Let
$\hat\rho$ be the density operator describing the signal state and let the
coherent amplitude of the LO be $\alpha=-A$ with $A$ real and positive. We now
prove the following Theorem:
\begin{theorem}
\label{theorem-wigner}
For $A$ very large (in the limit $A\to\infty$), which means a very strong LO,
the photon counting probability  $P^j_m$ is given by 
\begin{equation}
 P^j_m=\frac{{\rm e}^{-(2j-A^2)^2/2A^2}}
   {\sqrt{\pi}\,A^2} \,\langle x|\hat \rho |x\rangle,
\label{prob2}\end{equation}
where $|x\rangle$ is the eigenstate of the quadrature operator $\hat x=(\hat
a+\hat a^\dagger)/\sqrt 2$ with the eigenvalue $x=m/{\sqrt j}$.
\end{theorem}
{\em Proof:} We assume a pure signal state first,
$\hat\rho=|\psi\rangle\langle\psi|$ with
$\vert\psi\rangle=\sum_{n=0}^\infty\psi_n\,|n\rangle$.  The photon counting
probability $P^j_m$ is given by the square of the magnitude of the probability
amplitude $M^j_m$ for $j\pm m$ photons emerging from the first/second
interferometer output port,
\begin{equation}
M^j_m=\langle jm\vert\hat B(0,\pi/2,0)\vert-A\rangle_1 \vert\psi\rangle_2 ={\rm
          e}^{-A^2/2}\sum_{n=0}^{2j} \psi_n\,\frac{(-A)^{2j-n}}
          {\sqrt{(2j-n)!}}\,\,d^j_{m,j-n}(\pi/2).
\label{M2}\end{equation}
For $A\to\infty$, the probability distribution of the total photon number $2j$
is dominated by the Poissonian distribution of the photon number in the LO, so
$2j$ is sharply peaked at $A^2$.  Further, the photon number difference $2m$ at
the BS output is much less than $2j$ and also $n\ll2j$ holds for any photon
number $n$ for which $\psi_n$ is non-negligible.  This enables us to use
several approximations.  First, the fraction in Eq.~(\ref{M2}) can be
approximated via the Stirling formula and the Taylor expansion and by
neglecting terms of order $n^2/j$, $n/j$ and higher.  These approximations
yield
\begin{equation}
 \frac{(-A)^{2j-n}}{\sqrt{(2j-n)!}}
  \approx \frac{(-1)^{2j-n}}{\sqrt[4]{4\pi j}}\,
  {\rm e}^{A^2/2-(2j-A^2)^2/4A^2}.
\label{apr2}
\end{equation}
The condition $n\ll2j$ justifies the following asymptotic expression for
$d^j_{m,j-n}(\pi/2)$ that holds for $n\ll j$ \cite{Row01} and is central to the
calculation:
\begin{eqnarray}
d^j_{m,j-n}(\pi/2)&=&(-1)^n j^{-1/4} 
  \, u_n\left(\sqrt j\arcsin\frac{m}{j}\right)
        \nonumber       \\
        & \approx & (-1)^n j^{-1/4} \, u_n(m/\sqrt j).
\label{wigner}
\end{eqnarray}
Here $u_n(x)=\langle x\vert n\rangle$ denotes the $n^{\rm th}$ Hermite
Gaussian, that is, the $x$-representation of the number state $|n\rangle$. The
approximation $\sqrt j\arcsin(m/j)\approx m/\sqrt j$ is valid for $|m|\ll j$.
Substituting Eqs.~(\ref{apr2}) and (\ref{wigner}) into Eq.~(\ref{M2}) and
approximating $j$ by $A^2/2$ in the denominator, we obtain
\begin{equation}
M^j_m=\frac{{\rm e}^{-(2j-A^2)^2/4A^2}\,{\rm e}^{2{\rm i}\pi j}}
   {\sqrt[4]{\pi}\,A } \sum_{n=0}^{2j} \psi_n\,u_n(m/\sqrt j).
\label{M3}\end{equation}
The following identity for the inner product of the state $|\psi\rangle$ and
the eigenstate $|x\rangle$ holds due to completeness of the Fock basis:
\begin{equation}
  \langle x\vert\psi\rangle = \sum_{n=0}^{\infty}{}\langle x\vert n\rangle
   \langle n\vert\psi\rangle= \sum_{n=0}^{\infty} \,u_n(x)\,\psi_n.
\label{proj1}\end{equation}
Eq.~(\ref{proj1}) also holds if the summation over $n$ goes only to $2j$
instead of infinity because $2j\gg n$ for all $n$ for which $\psi_n$ differs
from zero significantly. Then Eq.~(\ref{M3}) becomes
\begin{equation}
M^j_m\doteq \frac{{\rm e}^{-(2j-A^2)^2/4A^2}\,
 {\rm e}^{2{\rm i}\pi j}}  {\sqrt[4]{\pi}\,A} \,
  \langle x|\psi\rangle
\label{homofinal}\end{equation}
with the eigenvalue $x=m/{\sqrt j}$. 
Eq.~(\ref{prob2}) is now obtained directly by squaring the magnitude of $M^j_m$
for the pure signal state.  The extension to mixed states is straightforward
and follows from linearity of quantum mechanics.~$\Box$

In the case of a general phase of LO when the amplitude is $\alpha=-A{\rm
e}^{\rm i\varphi}$, Eq.~(\ref{prob2}) turns into
\begin{equation}
 P^j_m=\frac{{\rm e}^{-(2j-A^2)^2/2A^2}}
   {\sqrt{\pi}\,A^2} \,\,{}_\varphi\langle x|\hat \rho |x\rangle_\varphi,
\label{prob3}\end{equation}
where $|x\rangle_\varphi$ is the eigenstate of the rotated quadrature $\hat
x_\varphi=\hat x\cos\varphi+\hat p\sin\varphi$ with the eigenvalue $x=m/\sqrt
j$.

From Theorem~\ref{theorem-wigner} we can now get the POVM defined by ${\rm
Tr}(\hat\rho\hat\Pi^j_m)=P^j_m$ and corresponding to BHD in the strong LO
limit:
\begin{equation}
 \hat\Pi^j_m=\frac{{\rm e}^{-(2j-A^2)^2/2A^2}}
   {\sqrt{\pi}\,A^2} \,|x\rangle\langle x|.
\label{POVM1}\end{equation}

Eqs.~(\ref{prob2}) and (\ref{POVM1}) show that in the limit of strong LO,
homodyne detection performs the POVM given by the projection $|x\rangle\langle
x|$ to the $x$--eigenstate. This fact has been known; however, here it has been
shown for the first time by a direct calculation. However, our result does not
provide any correction terms. We will obtain these in the next section by
employing the Glauber-Sudarshan P function. Before doing so, let us discuss a
few aspects of the result (\ref{prob2}).

First, the Gaussian factor in Eq.~(\ref{prob2}) reflects the fact that the
Poissonian distribution of the photon number for the LO (whence the
majority of the total $2j$ photons come) converges asymptotically to the
Gaussian distribution $P(2j)=(2\pi)^{-1/2}A^{-1}\exp[-(2j-A^2)/2A^2]$.

Second, one may wonder if the probability distribution (\ref{prob2}) is
properly normalized. Indeed, it is easy to check that
\begin{equation}
   \sum_{2j=0}^\infty\sum_{m=-j,-j+1,\dots,j}P^j_m=1
\end{equation}
by changing the double sum into an integral and using the normalization of the
state $\hat\rho$,
\begin{equation} 
  \int_{-\infty}^\infty  \langle x|\hat\rho|x\rangle\,{\rm d} x=1,
\end{equation}
and replacing $x=m/{\sqrt j}$ by $x=\sqrt2\,m/A$, which can be done for a
strong LO. 

Third, if the total photon sum $2j$ is not measured in the homodyne detection
scheme, then the probability distribution for the photon number difference $2m$
is
\begin{equation}
  P_m=\sum_{j=|m|}^\infty P^j_m
  =\frac1{\sqrt2\,A} \,\langle x|\hat\rho|x\rangle 
\label{noj}\end{equation}
(in the sum $j$ runs from $|m|$ to infinity via unit steps and the
eigenvalue $x$ is again $\sqrt2\,m/A$).  The factor $1/\sqrt2A$ in
Eq.~(\ref{noj}) is connected with the Jacobian $\sqrt2/A$ of the map $m\to
x=\sqrt2\,m/A$ and the fact that $m$ changes in half-integer steps.

\section{Glauber-Sudarshan $P$-Function Approach}

The method using the asymptotic formul{\ae} for SU(2) Wigner functions from the
previous section gave us the asymptotic expression for the photon counting
probability $P^j_m$. However, it is difficult to obtain the correction terms
due to the amplitude of the LO being finite because of absence of correction
terms in Eq.~(\ref{wigner}). This problem can be overcome by using the
Glauber-Sudarshan coherent-state representation, which we do in the following.

We represent the signal state $\hat\rho$ by the Glauber-Sudarshan
$P$-function \cite{Sud63,Gla63b,Kla68}
\begin{equation}
 \hat\rho=\int P(\beta)\,|\beta\rangle\langle\beta|\dd^2\beta.
\label{Prepres}\end{equation}
The BS input state is then 
\begin{equation}
\hat\rho_{\rm in}=|\alpha\rangle \langle\alpha|
  \otimes\int P(\beta)\,|\beta\rangle \langle\beta|
  \dd^2\beta,
\end{equation}
and the BS output state is
\begin{equation}
\hat\rho_{\rm out}
 = \int P(\beta)\,\left|\frac{\alpha-\beta}{\sqrt2}\right\rangle_1
                \left\langle\frac{\alpha-\beta}{\sqrt2}\right|    
      \otimes   \left|\frac{\alpha+\beta}{\sqrt2}\right\rangle_2
                \left\langle\frac{\alpha+\beta}{\sqrt2}\right| \dd^2\beta.
\end{equation}
Using Eq.~(\ref{1probab}), the probability $P^j_m$ is evaluated as
\begin{eqnarray}\nonumber
 P^j_m&=& e^{-|\alpha|^2}\int  P(\beta)\, e^{-|\beta|^2}\,\\
&=& \frac{2^{-2j}\,e^{-|\alpha|^2}}{(j+m)!\,(j-m)!}
 \int P(\beta)\,e^{-|\beta|^2}\,
 |\alpha-\beta|^{2(j+m)}\,|\alpha+\beta|^{2(j-m)}\dd^2\beta.
\label{matrixelement}
\end{eqnarray}
We again assume that the LO amplitude is $\alpha=-\A$.  Generalization to
arbitrary $\alpha$ is straightforward and discussed later. 

To evaluate the integral in Eq.~(\ref{matrixelement}), we use the following
identity, definitions and lemma that is proved in Appendix A:
\begin{identity}
For $|x|<1$ 
$$ (1+x)^n=\exp[n\ln(1+x)]
  =\exp\left[n\sum_{k=1}^\infty\frac{(-1)^{k-1}x^k}k\right]. $$
\label{expans}\end{identity}  

\begin{definition}
A pure {$z$--regular} state $|\psi\rangle$ is a state that can be expressed in
the Fock basis as
\begin{equation}
 |\psi\rangle={\cal N}\sum_{n=0}^\infty \frac{c_nz^n}{\sqrt{n!}}\,|n\rangle
\label{defzregular}\end{equation}
with the complex coefficients $c_n$ satisfying $|c_n|<1$, $z\in\mathbb R^+$ and
$\cal N$ a constant.  In other words, it is a state whose Fock basis
coefficients fall off at least as fast as those of a coherent state
$|z\rangle$.
\end{definition}

\begin{definition}
A mixed $z$--regular state is a finite mixture of pure $z$--regular states,
that is, a state corresponding to density operator
\begin{equation}
 \hat\rho=\sum_{i=1}^n p_i|\psi_i\rangle\langle\psi_i|
\end{equation}
with $n$ finite, $p_i\ge0$ and all $|\psi_i\rangle$ being $z$--regular.
\end{definition}

Examples of $z$--regular states include (i) a coherent state $|\gamma\rangle$
with $|\gamma|\le z$, (ii) superposition or mixture of several such coherent
states, (iii) superposition of such a coherent state with a number state, (iv)
superpositions or mixtures of several number states. However, they do not
include squeezed or thermal states.

\begin{lemma}
  The Glauber-Sudarshan P function $P(\beta)$ of a $z$--regular state is
  identically equal to zero for $|\beta|>z$.
\label{zeroP}\end{lemma}
(See Appendix A for the proof.)

We assume that the signal state is $z$-regular for some $z<A$.  Then
$P(\beta)=0$ for $|\beta|/A\ge1$, and we can employ Identity~\ref{expans}
in evaluating the powers $|\alpha+\beta|^{2(j+m)}$ and
$|\alpha-\beta|^{2(j-m)}$ in Eq.~(\ref{matrixelement}) as follows:
\begin{multline} 
 |\alpha+\beta|^{2(j+m)}\,|\alpha-\beta|^{2(j-m)}
 = A^{4j}
 \left(1+\frac{\beta}{\A}\right)^{j+m}\left(1+\frac{\beta^*}{\A}\right)^{j+m}
   \,\left(1-\frac{\beta}{\A}\right)^{j-m}
   \left(1-\frac{\beta^*}{\A}\right)^{j-m} \\
 =A^{4j}\,\left(
 2m\sum_{k=1}^\infty\frac1{2k-1}\,
 \frac{\beta^{2k-1}+(\beta^*)^{2k-1}}{\A^{2k-1}}
 -j\sum_{k=1}^\infty\frac1k\,\frac{\beta^{2k}+(\beta^*)^{2k}}{\A^{2k}}\right).
\label{expansion}\end{multline}
Using Eq.~(\ref{expansion}), the integral in Eq.~(\ref{matrixelement}) can
be expressed as
\begin{eqnarray} \nonumber
 I&\equiv& \int P(\beta)\,
 |\alpha-\beta|^{2(j+m)}\,|\alpha+\beta|^{2(j-m)}\dd^2\beta \\ 
 &=&A^{4j}e^{2m^2/\A^2} \int P(\beta)\,
 e^\chi\,e^{-2[(\beta+\beta^*)/2-m/\A]^2}\dd^2\beta,
\label{integral0}\end{eqnarray}
where the exponent
\begin{equation}
\chi=-\frac{2j-A^2}{2A^2}\{\beta^2+(\beta^*)^2\}
    +m\sum_{k=2}^\infty\frac2{2k-1}\,
 \frac{\beta^{2k-1}+(\beta^*)^{2k-1}}{\A^{2k-1}}
 -j\sum_{k=2}^\infty\frac1k\,\frac{\beta^{2k}+(\beta^*)^{2k}}{\A^{2k}}.
\label{E}\end{equation}
To evaluate the integral (\ref{integral0}), we will use the following
lemma.
\begin{lemma}\label{integralP}
Let $\hat\rho=\int P(\gamma)|\gamma\rangle\langle\gamma|\dd^2\gamma$ be the
Glauber--Sudarshan representation of the density operator $\hat\rho$.  Then for
$x\in\mathbb R$,
\begin{equation}
\int P(\gamma)\,
 \gamma^m(\gamma^*)^{n}e^{-[2^{-1/2}(\gamma+\gamma^*)-x]^2}\dd^2\gamma
={\sqrt\pi}\,{\rm Tr}\, \left(\hat\rho\,(\hat a^\dagger)^n
 |x\rangle\langle x|\,\hat a^m\right).
\label{Th1-1}\end{equation}
\end{lemma}
(For the proof see Appendix~\ref{lemma2}).

\begin{theorem}
For a $z$--regular state $\hat\rho$ and the LO coherent amplitude $-A$ with
$A>z$, 
\begin{multline}
P^j_m=
 \frac{\sqrt\pi\,2^{-2j}\,e^{-\A^2}\,A^{4j}\,e^{2m^2/\A^2}}{(j+m)!\,(j-m)!}
 \,\,{\rm Tr}\,\left\{\hat\rho\left[:\left|x =\frac{\sqrt2\,m}{A}
 \right\rangle\left\langle
 x=\frac{\sqrt2\,m}{A}\right|\right.\right.\\  \times    
 \exp\left(-\frac{2j-A^2}{2A^2}\{\hat a^2+(\hat a^\dagger)^2\}
 +2m\sum_{k=2}^\infty\frac1{2k-1}\,\frac{\hat a^{2k-1}
    +(\hat a^\dagger)^{2k-1}}{\A^{2k-1}} 
 \right. \\  \left.\left.\left.
 -j\sum_{k=2}^\infty\frac1k\,\frac{\hat a^{2k}
    +(\hat a^\dagger)^{2k}}{\A^{2k}}\right) :\right]\right\}.
\label{finalint}\end{multline}
The ordering symbol $:\,:$ that involves the projection operator
$|x\rangle\langle x|$ should be understood as
\begin{equation}
 :|x\rangle\langle x|\hat a^r(\hat a^\dagger)^s:
 \, =(\hat a^\dagger)^s|x\rangle\langle x|\hat a^r,
\label{ordering}\end{equation}
that is, all creation operators go to the left of the projector
$|x\rangle\langle x|$ and all annihilation operators go to the right of it.
\end{theorem}
{\em Proof:} The theorem is proved by a straightforward calculation applying
Lemma~\ref{integralP} to Eq.~(\ref{integral0}) and substituting the result into
Eq.~(\ref{matrixelement}).  $\Box$

The form of $P^j_m$ in Eq.~(\ref{finalint}) produces the POVM for
homodyne detection of a $z$-regular state:
\begin{multline} 
\hat\Pi^j_m=
 \frac{\sqrt\pi\,2^{-2j}\,e^{-\A^2}\,A^{4j}\,e^{2m^2/\A^2}}{(j+m)!\,(j-m)!}
\,\,\left\{:\left|x =\frac{\sqrt2\,m}{A}
 \right\rangle\left\langle
 x=\frac{\sqrt2\,m}{A}\right|\right. \\  \times
 \exp\left(-\frac{2j-A^2}{2A^2}\{\hat a^2+(\hat a^\dagger)^2\}
 +2m\sum_{k=2}^\infty\frac1{2k-1}\,\frac{\hat a^{2k-1}
    +(\hat a^\dagger)^{2k-1}}{\A^{2k-1}} \right. \\
 \left.\left. -j\sum_{k=2}^\infty\frac1k\,\frac{\hat a^{2k}
    +(\hat a^\dagger)^{2k}}{\A^{2k}}\right) :\right\},
\label{finalPOVM}\end{multline}
such that ${\rm Tr}\,\{\hat\rho\hat\Pi^j_m\}=P^j_m$ holds.

Eq.~(\ref{finalPOVM}) is the key result of our calculation. It shows that the
POVM for homodyne detection of a $z$-regular state (with $z<A$) is given by the
normally ordered product of the projector $|x\rangle\langle x|$ multiplied by
an exponential of powers of creation and annihilation operators.  We will
discuss this result in the following sections.

We still need to mention the case of a general phase of the LO when
$\alpha=-A{\rm e}^{{\rm i}\varphi}$. The operators $\hat a$ and $\hat
a^\dagger$ in Eqs.~(\ref{finalint}) and (\ref{finalPOVM}) then have to be
replaced by $\hat a_\varphi={\rm e}^{{\rm i}\varphi}\hat a$, $\hat
a_\varphi^\dagger={\rm e}^{-{\rm i}\varphi}\hat a^\dagger$, respectively, and
$|x\rangle$ has to be replaced by $|x\rangle_\varphi$, the eigenstate of the
rotated quadrature $\hat x_\varphi=\hat x\cos\varphi+\hat p\sin\varphi=(\hat
a_\varphi+\hat a_\varphi^\dagger)/{\sqrt2}$ with the eigenvalue
$x=\sqrt2\,m/A$.

\subsection{Limit $A\to\infty$}

We begin discussing the result (\ref{finalint}) by considering the limit of
strong LO, that is, the limit $A\to\infty$ for a given signal state $\hat\rho$.
This will give us the asymptotic expression for the photon counting probability
$P^j_m$ corresponding to an ideal homodyne detection.

For large $A$, the total photon number distribution is dominated by the
Poissonian LO distribution, so $2j$ is peaked at $A^2$ and has the variance of
$A^2$. Hence the expression $(2j-A^2)/2A^2$ in the exponent of
Eq.~(\ref{finalint}) is negligible.  At the same time, in the sums in the
exponent the factors $A^{-k}$ go to zero for $A\to\infty$.  Thus the trace in
Eq.~(\ref{finalint}) becomes simply ${\rm Tr}(\hat\rho|x\rangle\langle
x|)=\langle x|\hat\rho|x\rangle$.  The factor in front of the trace can be
approximated using the Stirling formula for the factorials and neglecting terms
of order $(2j-A^2)/2A^2$ and $m/A^{3/2}$.  Then the probability $P^j_m$
becomes
\begin{equation} 
  P^j_m=\frac{{\rm e}^{-(2j-A^2)^2/2A^2}}{\sqrt\pi\,A^2}
   \,\langle x|\hat\rho|x\rangle,
\label{probab_A_to_infty}\end{equation}
which replicates the result~(\ref{prob2}) from the Sec.~\ref{asymp}. The only
difference is that in Eq.~(\ref{prob2}) the eigenvalue was $x=m/\sqrt j$ while
here we have $x=\sqrt2\,m/A$.  However, this difference is not important as $j$
is sharply peaked about $A^2/2$ for a strong LO as has been mentioned.

\subsection{Infinite series and its convergence}

For a finite amplitude of LO, one can expand the exponential function in
Eq.~(\ref{finalint}) using the usual Taylor series. This gives an expansion of
the the photon counting probability $P^j_m$ into the following series:
\begin{multline} 
P_m^j=
 \frac{\sqrt\pi\,2^{-2j}\,e^{-\A^2}\,A^{4j}\,e^{2m^2/\A^2}}{(j+m)!\,(j-m)!}
 \,\,\biggl\{\langle x|\hat\rho|x\rangle
  -\frac{2j-A^2}{2A^2}[\langle x|\hat a^2\hat\rho|x\rangle+\langle x|\hat\rho
  (\hat a^\dagger)^2|x\rangle]\\
 +\frac{2m}{3\A^3}[\langle x|\hat a^3\hat\rho|x\rangle+\langle x|\hat\rho
  (\hat a^\dagger)^3|x\rangle]+\dots \biggr\}.
\label{truncatedint}\end{multline}
The terms in the series are arranged such as to contain increasing powers of
creation and annihilation operators.  To determine for which states this series
converges is a task that we have not been able to solve in general. We believe,
though, that the following conjecture is valid:
\begin{conjecture}
The series in Eq.~(\ref{truncatedint}) converges for all $z$-regular states
with $z<A$.
\end{conjecture}
Surprisingly enough, however, it turns out that the question of convergence
does not really matter for practical purposes as we will see in the following
section.

In addition, also the factor in front of the parentheses in
Eqs.~(\ref{finalint}) or (\ref{truncatedint}) can be expanded into a series
using the Stirling formula for the factorials and Taylor expansion around the
point $m/j=0$ and $(2j-A^2)/2A^2=0$. The leading term of the series for this
factor is the same fraction as in Eq.~(\ref{probab_A_to_infty}) and reflects
the Gaussian limit of the Poissonian distribution for the LO photon number. We
do not write the other terms explicitly.

\subsection{Truncation in Fock basis}

We explore the properties of the series (\ref{truncatedint}) for density
operators truncated in the Fock basis.  Such density operators can be expressed
as
\begin{equation}
 \hat\rho=\sum_{m,n=0}^N\rho_{mn}\,|m\rangle\langle n|
\label{rhotruncated}\end{equation}
for some finite $N$.

\begin{theorem}
For truncated signal states $\hat\rho$, the series (\ref{truncatedint}) is
finite (i.e., it contains only a finite number of non-zero terms). Therefore it
converges and expresses the exact photon counting probability $P_m^j$.
\end{theorem}
{\em Proof:} Consider a term in the series in Eq.~(\ref{truncatedint}) that
contains more than $2N$ field operators (i.e., annihilation and creation
operators).  Then it contains more than $N$ creation and/or more than $N$
annihilation operators.  As all the annihilation operators are to the left from
the density operator $\hat\rho$ and all creation operators are to the right of
it, every such term turns into zero because of the
truncation~(\ref{rhotruncated}) of $\hat\rho$. Further, it follows from the
expansion of an exponential in Eq.~(\ref{finalint}) that in the series in
Eq.~(\ref{truncatedint}) the number of terms with less than $k$ field operators
is finite for every $k$.  Hence, the number of nonzero terms in the series in
Eq.~(\ref{truncatedint}) is finite, which we wanted to prove.  $\Box$

The fact that the series converges for truncated states is very useful as it
can be employed for states for which the series does not converge. The reason
is the following. Consider a general state
$\hat\rho=\sum_{m,n=0}^\infty\rho_{mn}\,|m\rangle\langle n|$ and for a given
cutoff $N\in{\mathbb N}$ define the corresponding truncated state $\hat\rho'$
with matrix elements $\rho'_{mn}$ satisfying
\begin{equation}
 \rho'_{mn}=
\begin{cases} 
 \bigl({\sum_{i=1}^N\rho_{ii}}\bigr)^{-1} \,\rho_{mn}&\quad 
  {\rm for}\,\, m\le N,n\le N \cr
   0 &\quad {\rm otherwise}
\end{cases}
\end{equation}
This definition ensures the proper normalization of $\hat\rho'$.  Now, the
cutoff number $N$ can be chosen arbitrarily large, so that the truncated state
$\hat\rho'$ mimics the state $\hat\rho$ arbitrarily close. Then also the photon
counting probabilities $P_m^j{}'$ corresponding to the state $\hat\rho'$ can be
brought arbitrarily close to the probabilities $P_m^j$ for all pairs of $j,m$,
for which $P_m^j$ is non-negligible. This enables us to employ
Eq.~(\ref{truncatedint}) for calculating $P_m^j$ with an arbitrary precision
also for states, for which the series (\ref{truncatedint}) does not even
converge.

Another question concerns the practical usefulness of this truncation
procedure. To see an example when it is not useful, consider the signal state
as a coherent state with an amplitude $\beta, |\beta|\gg A$, and its truncation
for a very large $N$ (say $N\gg|\beta|^2$). In this situation, the series
(\ref{truncatedint}) diverges while after the truncation it becomes finite and
so it converges.  A closer inspection of Eq.~(\ref{truncatedint}) also shows
that the initial subsequent terms grow very quickly for both the original and
truncated states. Therefore we would need very many of them to calculate the
probabilities $P^j_m$ using the truncation procedure and
Eq.~(\ref{truncatedint}), which would not be very practical and it would be
much simpler to calculate $P_m^j$ directly. This can be expected 
as the signal field is not weaker than the LO field. 

On the other hand, in many situations our result is very useful.  Our
calculations were motivated by trying to show that homodyne detection measures
the field quadrature, and to find correction terms. This happens for large
amplitudes of LO when the term $\langle x|\hat\rho|x\rangle$ in the photon
counting probability $P_m^j$ is the largest and dominant one. In such
situations the truncation works very well and the series (\ref{truncatedint})
gives good correction terms for balanced homodyne detection as can be seen in
the following section.

We should also note that the convergence of the series (\ref{truncatedint}) is
not directly related to the behavior of the initial terms.  It can happen
(e.g. for a weak thermal state or a weakly squeezed vacuum state) that the
initial subsequent terms decrease quickly but after some time, they start to
grow and the series diverges. At the same time, for weak signal states
(compared to the LO) these first terms provide an increasingly good
approximation to the photon counting probability $P_m^j$ as can be seen in the
next section with numerical simulations.  The situation is thus similar to the
one in perturbation theory: even though a perturbation series diverges, its
several (or many) initial terms may give a good approximation.

\subsection{What is a strong local oscillator?}

We would like to address the question now of when the LO is strong enough so
that BHD really performs the projective measurement of the quadrature phase of
the signal field. It can be roughly said that it is in situations for which the
first term $\langle x|\hat\rho|x\rangle$ in the brackets in
Eq.~(\ref{truncatedint}) dominates over the remaining ones. Let us focus at the
second and third terms,
\begin{equation}
  \frac{2j-A^2}{2A^2}[\langle x|\hat a^2\hat\rho|x\rangle+\langle x|\hat\rho
  (\hat a^\dagger)^2|x\rangle],
\label{second_term}\end{equation}
and try to estimate their magnitude compared to $\langle
x|\hat\rho|x\rangle$. First, the distribution of the LO photon number is
Poissonian with both mean and variance equal to $A^2$. Therefore, if we assume
that the LO contains many more photons than the signal state, the quantity
$(2j-A^2)/2A^2$ is of order of $1/A$. Of course, $2j$ can be an arbitrary
integer, but if it is not close enough to $A^2$, the probability $P^j_m$
becomes negligible. In this sense we mean that $(2j-A^2)/2A^2$ is of order of
$1/A$.

To estimate $\langle x|\hat a^2\hat\rho|x\rangle+{\rm c.c.}$, we will consider
two different types of signal states -- a coherent state and a number state.
The discussion for a general state would be very difficult, and we think that
coherent and number states are good representatives that can help us
understand the general behavior of the series in Eq.~(\ref{truncatedint}).

For a coherent state $|\beta\rangle$ for which
$\hat\rho=|\beta\rangle\langle\beta|$,
\begin{equation}
\langle x|\hat a^2\hat\rho|x\rangle+\langle x|\hat\rho (\hat
a^\dagger)^2|x\rangle= 2\,{\rm Re}\{\beta^2\}\,\langle x|\hat\rho|x\rangle.
 \end{equation}
This means that the term~(\ref{second_term}) in the series~(\ref{truncatedint})
is of order of ${\rm Re}\{\beta^2\}/A$ compared to the first term $\langle
x|\hat\rho|x\rangle$. We see that if the mean photon number in the signal state
is much less than the magnitude of the LO amplitude, the leading term is
dominant. 

For the signal field in a number state $|n\rangle$ we have 
\begin{equation}
\langle x|\hat a^2\hat\rho|x\rangle+\langle x|\hat\rho (\hat
a^\dagger)^2|x\rangle= 2\sqrt{n(n-1)}\,{\rm Re}\{\langle x|n\rangle\langle
x|n-2\rangle^*\}.
\end{equation}
The magnitude of the inner product $\langle x|n-2\rangle^*$ can be considered
roughly the same as that of $\langle x|n\rangle^*$ for our purpose.  As
$\sqrt{n(n-1)}$ is close to $n$ for $n>1$, we arrive at a similar result as for
the coherent state: the second and third terms become unimportant if $A$ is
much larger than the photon number in the signal state.

The analysis of the magnitude of other terms in Eq.~(\ref{truncatedint}) would
be similar.  The result is that if $A\gg\overline n$, where $\overline n$ means
the average photon number in the signal state, the subsequent terms decrease
quickly and homodyne detection indeed measures the field quadrature phase. It
should be noted that it is not enough if the mean number of photons $\overline
n$ in the signal state is much less than the number of photons in the LO; in
fact, the correct condition is that the {\em square} of $\overline n$ must be
much smaller than the number of photons in the LO.

This condition has a clear physical interpretation. As the photon number in the
coherent state $|-A\rangle$ has a Poissonian distribution of width $A$, the
condition of a strong LO can be formulated such that the mean photon number in
the signal must be much less than the {\em fluctuation} of the photon number in
LO. Now suppose for a moment that the opposite would hold. Then from the
knowledge of the total photon number $2j$ we could access some information
about the photon number in the signal state. However, the photon number
operator does not commute with the quadrature $\hat x$, so this would
necessarily disturb the measurement of $\hat x$. On the other hand, if the
strong LO condition is satisfied, then we do not know how many of the $2j$ 
photons come from the signal and how many come from the LO; thus, the
different possibilities can interfere and the distribution of $\hat x$ is not
affected.

\subsection{Numerical simulations}
  
In this section we show some numerical simulations of our results.  For a given
pure signal state $|\psi\rangle$ and a given photon number sum $2j$ we compare
the exact photon counting probability $P^j_m$ calculated with the help of
Eq.~(\ref{M2}) with the series~(\ref{truncatedint}) truncated at different
points.  The purpose of such a simulation is to show that taking increasing
number of terms in the series~(\ref{truncatedint}) gives an increasingly better
approximation to the exact probability $P_m^{j}$.

The LO amplitude was chosen to be $\alpha=-A=-20$ which means that the mean
photon number of the LO field is 400. The value of $j$ in the individual plots
was chosen randomly from the Poissonian distribution of LO photon number. It
has turned out during the simulations that changing $j$ inside the interval for
which the probability $P^j_m$ is non-negligible does not affect the behavior
of the series significantly.  As the signal states we have chosen a coherent
state with amplitude $2$, a squeezed vacuum state~\cite{Rad87} $\exp[r(\hat
a^2-\hat a^{\dagger2})/2]\,|0\rangle$ with $r=1.5$ and a number state
$|6\rangle$.  The results of the simulations are shown in Fig.~\ref{simul}.
The exact probabilities $P^j_m$ are shown in black, and the results of
truncation of the series~(\ref{truncatedint}) keeping terms with (i)~zero
number of field operators,
\begin{equation}
P_m^{j\,(0)}=
\frac{\sqrt\pi\,2^{-2j}\,e^{-\A^2}\,A^{4j}\,e^{2m^2/\A^2}}{(j+m)!\,(j-m)!}
\, \langle x|\hat\rho|x\rangle,
\end{equation}
are shown in green color, (ii)~maximum of two field operators 
\begin{equation}
P_m^{j\,(2)}=
\frac{\sqrt\pi\,2^{-2j}\,e^{-\A^2}\,A^{4j}\,e^{2m^2/\A^2}}{(j+m)!\,(j-m)!}
\, \left\{\langle x|\hat\rho|x\rangle-\frac{2j-A^2}{2A^2}
[\langle x|\hat a^2\hat\rho|x\rangle+\langle x|\hat\rho
  (\hat a^\dagger)^2|x\rangle]\right\}
\end{equation}
are shown in blue color, and (iii)~maximum of four field operators are shown in
red (we do not write $P_m^{j\,(4)}$ explicitly).

The simulations show that with increasing number of terms in the
series~(\ref{truncatedint}), a better approximation to the exact photon
counting probability is achieved.

\section{Conclusion}
    
We have analyzed balanced homodyne detection in terms of the POVM for photon
counting by directly calculating the photon counting probability.  We employed
two different approaches. First, using asymptotic expressions for SU(2) Wigner
functions allowed us to establish the non-trivial connection between the
discrete variables $j,m$ corresponding to photon numbers being detected and the
continuous quadrature phase variable $x_\varphi$.  In the strong LO limit, we
have shown that homodyne detection indeed performs the projective measurements
corresponding to POVM $|x\rangle\langle x|$, where $|x\rangle$ is the
eigenstate of quadrature phase operator.  Second, employing the
Glauber-Sudarshan $P$-function, we extended the result obtained by the first
approach. For a very large amplitude of the LO, the result was the same, and
for finite amplitudes we obtained additional correction terms.  Even though the
series we got does not converge in general, it can be used for determining the
correction terms via truncation of the signal state in the Fock basis.  We have
determined the strong LO condition for coherent and number signal states -- the
square of the mean photon number in the signal state must be much smaller than
the mean photon number in the LO.  We have also performed numerical simulations
that confirm the validity of the quadrature-phase POVM and the correction terms
for a LO that is not strong for typical signal states. Therefore, in addition
to obtaining the quadrature-phase POVM rigorously from the photon counting
POVM, we have an expansion that yields correction terms for the POVM that works
well for typical signal states in quantum optics.

In this paper we have considered a perfect HD scheme with ideal detectors, LO
and BS and 100\% mode-matching. In practice, all these elements are subject
to imperfections, which disturbs the measurement. 
For example, the LO from a realistic laser has an amplitude distribution
$P_{\rm LO}(\alpha)$ broader than the delta-function. This would convolute
the probability $P^j_m(\alpha)$, where we now write the dependence on $\alpha$
explicitly, with $P_{\rm LO}(\alpha)$. If this distribution is Gaussian, then
the strong LO measurement would correspond to a Gaussian spread of the
quadrature measurement with the imprecision corresponding to the degree of 
LO amplitude fluctuation. 
Lossy beamsplitters and inefficient photodetectors would add vacuum noise that
would result in Gaussian spread of quadrature measurement, similar to the
effect discussed above for the LO amplitude spread.
Finally, for a multi-mode field with the LO mode-matching condition satisfied,
the detection  efficiency can incorporate the mismatch between the beam and
detector modes. 
If, on the other hand, the signal and LO modes are mismatched, HD efficiency
declines, and beats between different frequency modes arise. 

Our operational approach to HD ignores the realistic effects described above,
but the theory is readily generalized to accommodate these effects by including
inefficiencies and multimode description. Moreover, if multimode fields and
beats are desirable, heterodyne detection replaces homodyne detection (for
which signal and LO are frequency matched); an operational formulation of
heterodyne detection without mean field approximation is a topic of further
research.

\acknowledgments

We appreciate valuable discussions with V.~Bu\v zek, S.~Bartlett, H.\ de Guise,
T.~Rudolph, Ch.~Simon and M.~Lenc plus support by Macquarie University Research
Grants and an~Australian Research Council Large Grant.  We acknowledge the
support of the Erwin Schr\"odinger International Institute for Mathematical
Physics in Vienna during the early stages of this project.  BCS acknowledges
support from the Quantum Entanglement Project ICORP, IST at the Ginzton
laboratory during certain stages of this work, and support from Alberta's
informatics Circle of Research Excellence (iCORE). TT acknowledges a kind
hospitality of Department of Physics, Macquarie University Sydney.

\appendix
\section{Properties of the P representation for $z$--regular states}

We first prove Lemma~\ref{zeroP} for pure $z$--regular states and then
generalize to mixed states.  The density operator associated with a normalized
pure state $|\psi\rangle$ is
\begin{equation}
  \hat \rho=|\psi\rangle\langle\psi|,
\end{equation}
and is represented by the Glauber-Sudarshan P representation according to
Eq.~(\ref{Prepres}). The trace of $\hat\rho$ is unity due to the normalization
of the state $|\psi\rangle$ and hence the integral of the $P$--function over
the complex plane is equal to unity:
\begin{equation}
 \int P(\beta)\,{\rm d}^2\beta=\int P(\beta){\rm
 Tr}\,|\beta\rangle\langle\beta|\,{\rm d}^2\beta={\rm Tr}\, \hat\rho=1.
\end{equation}
Now, for a positive number $r$ we define a non-unitary operator
\begin{equation}
  \hat S(r)={\rm e}^{-r}\,\exp\left(\frac{r\hat a}z\right)
\end{equation}
and, for a normalized $z$-regular state $|\psi\rangle$ [see
Eq.~(\ref{defzregular})], we consider the state
\begin{equation}
 |\psi'\rangle=\hat S(r)|\psi\rangle
  ={\cal N}\sum_{n=0}^\infty \frac{c'_{n}z^n}{\sqrt{n!}}\,|n\rangle,
\label{psi'}\end{equation}
where the  coefficients $c'_n$ are related to the coefficients
$c_n$ by
\begin{equation}
 c_n'={\rm e}^{-r}\,\sum_{m=0}^\infty \frac{c_{n+m}r^m}{m!}
 =\frac{\sum_{m=0}^\infty c_{n+m}r^m/m!}{\sum_{m=0}^\infty r^m/m!} .
\end{equation}
Clearly $|c'_n|\le 1$, so the state $|\psi'\rangle$ is also $z$ regular but
generally not normalized. To normalize it, we introduce the inverse norm ${\cal
N'}=\langle\psi'|\psi'\rangle^{-1/2}$ so that the state $ |\psi'\rangle_N={\cal
N'} |\psi'\rangle =\langle\psi'|\psi'\rangle^{-1/2} |\psi'\rangle $ is
normalized.  The density operator $\hat\rho'$ of $|\psi'\rangle_N$ can be
expressed via the density operator $\hat\rho$ as
\begin{equation}
 \hat\rho'={\cal N'}^2
 {\rm e}^{-2r}\,\exp\left(\frac{r\hat a}z\right)
 \hat\rho\exp\left(\frac{r\hat a^\dagger}z\right)
\label{rho'}\end{equation}
and the P function corresponding to $|\psi'\rangle_N$ is hence
\begin{equation}
 P'(\beta)={\cal N'}^2
 {\rm e}^{-2r}\,\exp\left(\frac{r\beta}z\right)
  \exp\left(\frac{r\beta^*}z\right)\,P(\beta) ={\cal N'}^2
   \exp\left[2r\left(\frac{{\rm Re\,}\beta}z-1\right)\right]\,P(\beta).
\label{P'}\end{equation}
The integral of $P'$ over the complex plane is unity as the state
$|\psi'\rangle_N$ is normalized:
\begin{equation}
  \int P'(\beta)\dd^2\beta= 1
\label{intP'}\end{equation}
Decomposing $\beta$ to real and imaginary parts $\beta=\beta_1+\ii\beta_2$,
and using Eq.~(\ref{P'}), we can write Eq.~(\ref{intP'}) as a double integral
\begin{eqnarray}\nonumber
 1&=&{\cal N'}^2
 \,\int_{-\infty}^\infty\dd\beta_1\,\exp\left[2r\left(
  \frac{\beta_1}z-1\right)\right]\int_{-\infty}^\infty\dd\beta_2
  P(\beta_1,\beta_2)\\
  &=&{\cal N'}^2
  \,\int_{-\infty}^\infty \exp\left[2r\left(
  \frac{\beta_1}z-1\right)\right]\,G(\beta_1) \dd\beta_1,
\label{PG}\end{eqnarray}
where we have denoted
\begin{equation}
 G(\beta_1)=\int_{-\infty}^\infty P(\beta_1,\beta_2)\dd\beta_2.
\end{equation}
The inner product $\langle\psi'|\psi'\rangle$ can be bound as follows [see
Eq.~(\ref{psi'})]:
\begin{equation}
 \langle\psi'|\psi'\rangle={\cal N'}^{-2}
 ={\cal N}^2\sum_{n=0}^\infty
 \frac{|c'_{n}|^2z^{2n}}{n!}\le{\cal N}^2\sum_{n=0}^\infty
 \frac{z^{2n}}{n!}={\cal N}^2{\rm e}^{z^2} .
\label{product}\end{equation}
From Eqs.~(\ref{PG}) and (\ref{product}),
\begin{equation}
 \int_{-\infty}^\infty
 \exp\left[2r\left(\frac{\beta_1}z-1\right)\right]\,G(\beta_1)
 \dd\beta_1\le {\cal N}^2{\rm e}^{z^2} . 
\label{int}\end{equation}
We see that the integral (\ref{int}) is bound by a fixed number ${\cal N}^2{\rm
e}^{z^2}$, no matter how large $r$ we choose. The only way to satisfy this is
if $G(\beta_1)\equiv0\,\, \forall\beta_1>z$. Specifically if $f(x):\mathbb
R\to\mathbb R$ is a function and we know that
\begin{equation}
  \int_{-\infty}^\infty f(x)\,{\rm e}^{ax}\dd x < c
\end{equation}
where $c>0$ is fixed and $a$ is arbitrary positive, then necessarily
$f(x)=0$ for all $x>0$.
Thus, we obtain
\begin{equation}
  G(\beta_1)=\int_{-\infty}^\infty P(\beta_1,\beta_2)\dd\beta_2=0\quad{\rm
  for} \quad \beta_1>z,
\end{equation}
which means that the integral of $P$ over any vertical line in the
complex plane that is farther than $z$ from the origin is zero.

\begin{figure}[h]
\mbox{\epsfxsize=50mm\epsfbox{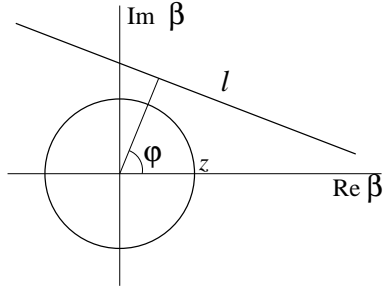}}
\caption{The integral of the P function of the state $|\psi\rangle$ over the
  line $l$ shown in the picture vanishes, and so it does for all other lines
  that do not intersect the circle $|\beta|=z$.  Therefore the P function
  itself vanishes outside the circle, i.e., for $|\beta|>z$.}
\label{line}
\end{figure}

Now the whole construction can be repeated with another state
\begin{equation}
 |\psi'_\varphi\rangle
 ={\rm e}^{-r}\,\exp\left(\frac{re^{\ii\varphi}\hat a}z\right)|\psi\rangle
\label{psi'phi}\end{equation}
whose P function is
\begin{equation}
 P'_\varphi(\beta) =\langle\psi'_\varphi|\psi'_\varphi\rangle^{-1}
 \exp\left[2r\left(\frac{{\rm Re\,}
      (\beta e^{\ii\varphi})}z-1\right)\right]\,P(\beta).
\end{equation}
Using the same argument, we arrive at the fact that the integral of $P(\beta)$
over any line whose normal has the angle $\varphi$ with the real axis and whose
distance from the origin is larger than $z$ (eg the line $l$ in
Fig.~\ref{line}) is zero. Now, as $\varphi$ can be arbitrary, this means that
the integral over {\em all} lines not intersecting the circle with radius $z$
is zero. Then it follows by the tomographic argument that the P-function must
be zero outside the circle, which is what we wanted to prove.

The generalization of the claim to mixed $z$ regular states is straightforward
as the P-function of a mixed state is the weighed sum of the P-functions of the
pure states in the mixture.

\section{Proof of Lemma~\ref{integralP}}
\label{lemma2}

A direct calculation yields
\begin{eqnarray}\nonumber
{\rm Tr}\,\left(\hat\rho\,(\hat a^\dagger)^n
 |x\rangle\langle x|\,\hat a^m\right)\hat\rho
&=& \int P(\gamma)\,\langle x|\,
     \hat a^m |\gamma\rangle\langle\gamma|
 (\hat a^\dagger)^n\,|x\rangle\dd^2\gamma\\\nonumber
&=&\int P(\gamma)\,\gamma^m(\gamma^*)^{n}
  |\langle x|\gamma\rangle|^2\dd^2\gamma \\
&=&\frac1{\sqrt\pi}\int P(\gamma)\,
 \gamma^m(\gamma^*)^{n}e^{-[2^{-1/2}(\gamma+\gamma^*)-x]^2}\dd^2\gamma.
\end{eqnarray}
Here the fact that $|\langle x|\gamma\rangle|^2=\pi^{-1/2}
\exp[-(x-\sqrt2\,{\rm Re\{\gamma\})^2]}$ was used.


\begin{figure}[h]
\includegraphics*[width=10cm,keepaspectratio]{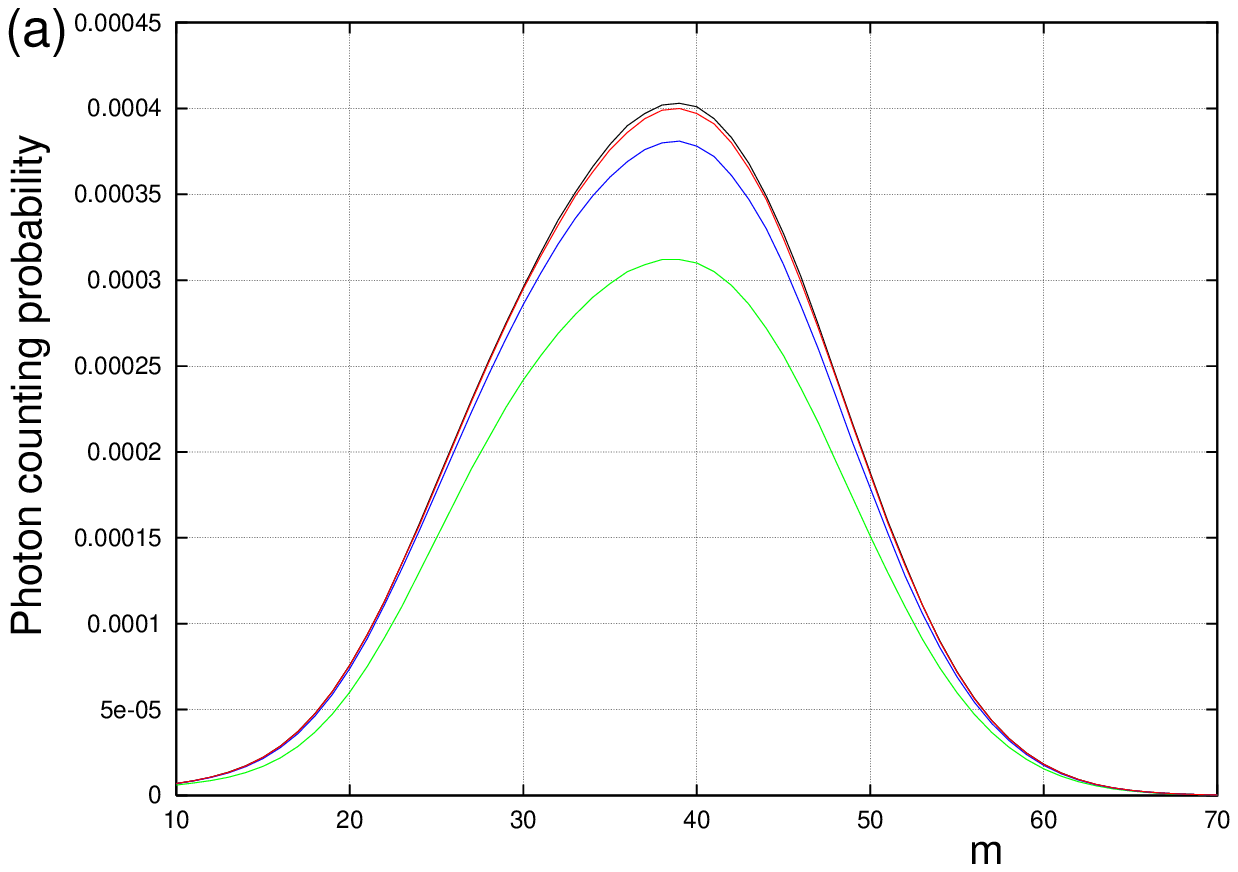} \\
\includegraphics*[width=10cm,keepaspectratio]{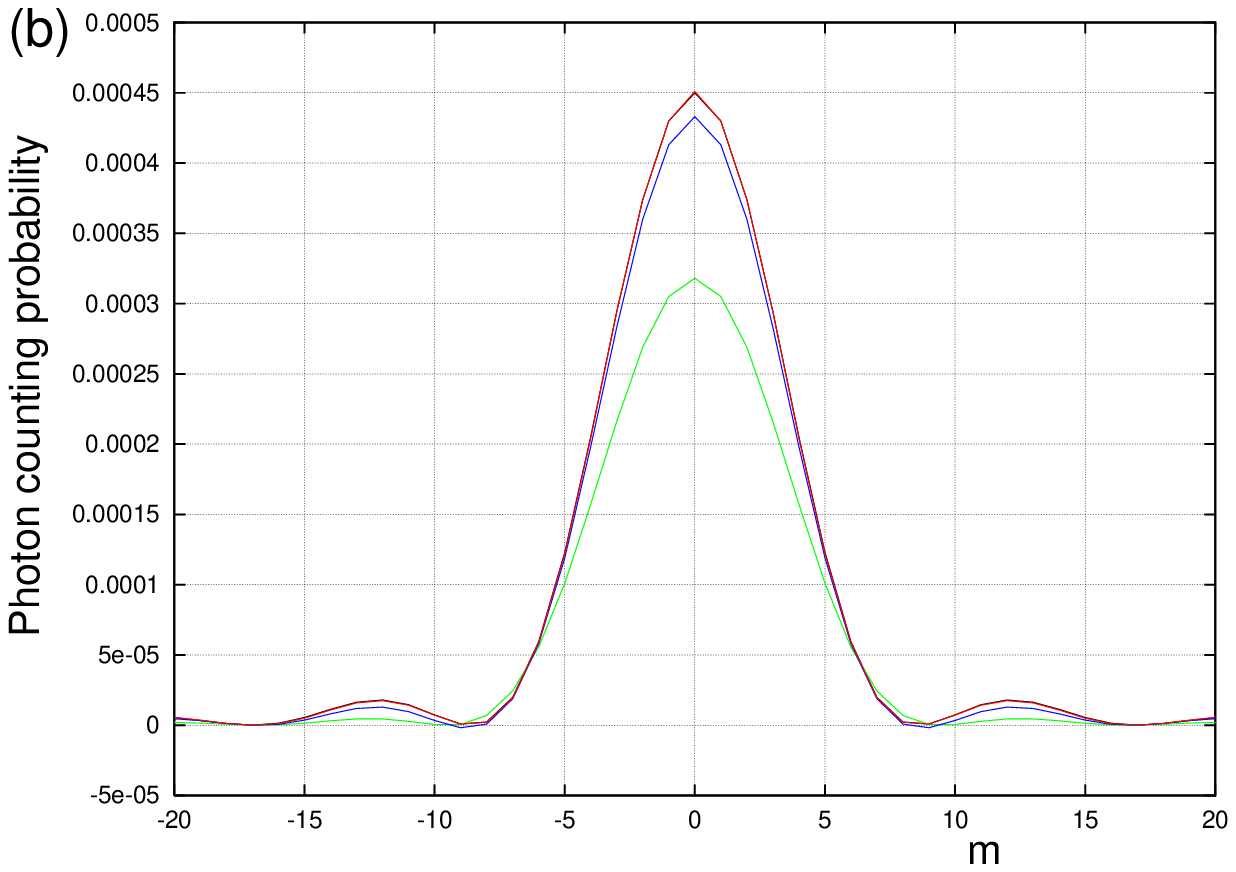} \\
\includegraphics*[width=10cm,keepaspectratio]{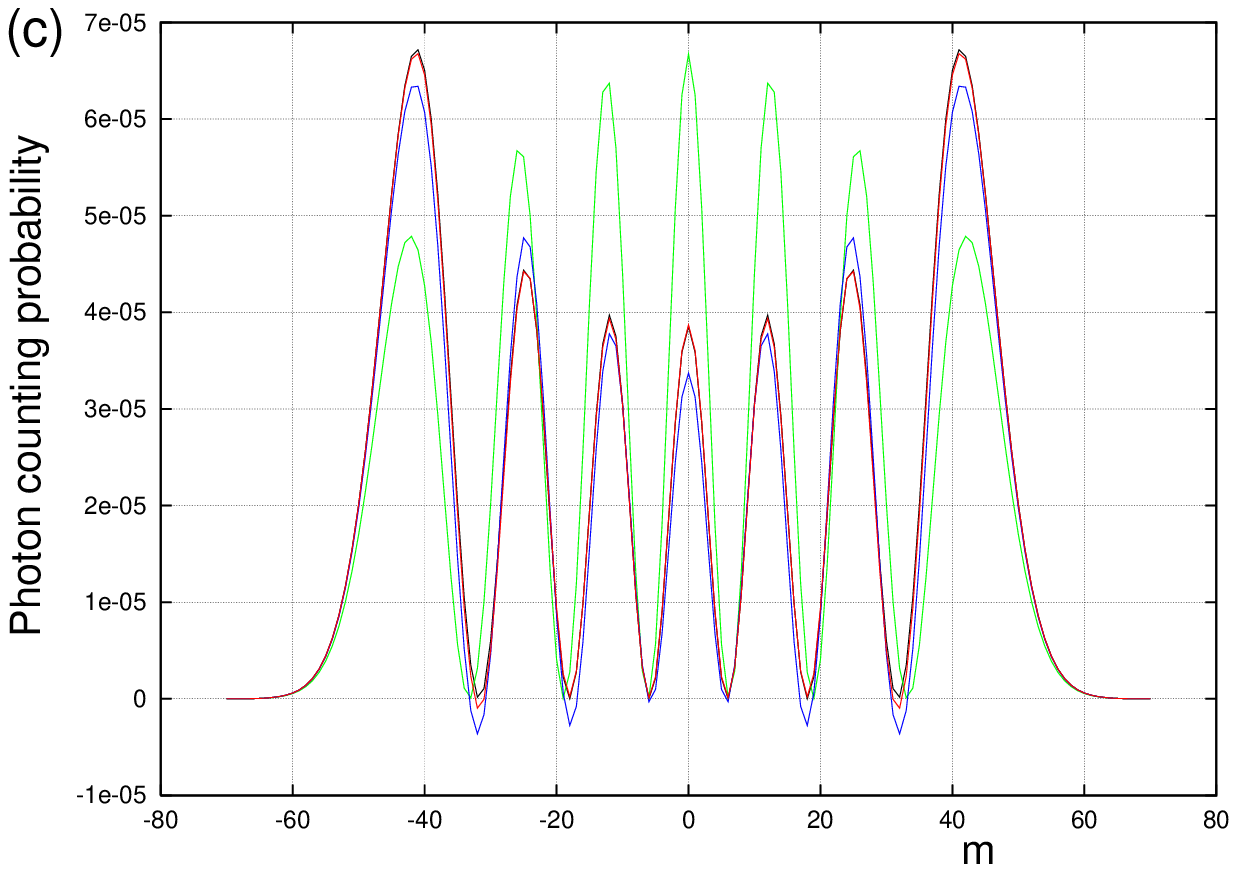} \\
\caption{The simulation for (a) coherent state $|\gamma\rangle$ with $\gamma=2$
 for $j=190$, (b) squeezed state with squeezing parameter $r=1.5$ for
 $j=219.5$, and (c) number state $|6\rangle$ for $j=183.5$. The exact
 probabilities are shown in black, and the truncated ones are shown in green,
 blue and red, respectively, according to the increasing number of terms in
 Eq.~(\ref{truncatedint}) taken into account. The red curves are so close to
 the black ones in (b) and (c) that they almost cover them in the plots.}
\label{simul}
\end{figure}

\end{document}